%% file: cscw24.tex
\documentclass[acmsmall]{acmart}

\setcopyright{rightsretained}
\acmJournal{PACMHCI}
\acmYear{2024} \acmVolume{8} \acmNumber{CSCW1} \acmArticle{92} \acmMonth{4}\acmDOI{10.1145/3637369}

\makeatletter
\gdef\@copyrightpermission{
  \begin{minipage}{0.2\columnwidth}
   {\includegraphics[width=0.90\textwidth]{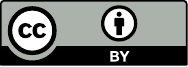}}
  \end{minipage}\hfill
  \begin{minipage}{0.8\columnwidth}
   \href{https://creativecommons.org/licenses/by/4.0/}{This work is licensed under a Creative Commons Attribution International 4.0 License.}
  \end{minipage}
  \vspace{5pt}
}
\makeatother
\usepackage{algorithm,algpseudocode}
\usepackage{ulem}
\usepackage{subfigure}

\begin{document}

\title{Mapping the Landscape of Independent Food Delivery Platforms in the United States}

\author{Yuhan Liu}
\authornote{Both authors contributed equally to this research.}
\email{yl8744@princeton.edu}
\orcid{0000-0001-6852-6218}
\affiliation{%
  \institution{Princeton University}
  \country{USA}
}

\author{Amna Liaqat}
\authornotemark[1]
\email{al0910@princeton.edu}
\orcid{0000-0002-5170-1945}
\affiliation{%
  \institution{Princeton University}
  \country{USA}
}

\author{Owen Xingjian Zhang}
\email{owenz@princeton.edu}
\orcid{0009-0008-2949-7379}
\affiliation{%
  \institution{Princeton University}
  \country{USA}
}

\author{Mariana Consuelo Fernández Espinosa}
\email{mferna23@nd.edu}
\orcid{0009-0004-1116-2002}
\affiliation{%
  \institution{University of Notre Dame}
  \country{USA}
}

\author{Ankhitha Manjunatha}
\email{am5897@princeton.edu}
\orcid{0009-0006-4341-6948}
\affiliation{%
  \institution{Princeton University}
  \country{USA}
}

\author{Alexander Yang}
\email{ayang130@umd.edu}
\orcid{0009-0000-4212-0228}
\affiliation{%
  \institution{University of Maryland}
  \country{USA}
}
\author{Orestis Papakyriakopoulos}
\email{orestis.p@tum.de}
\orcid{0000-0003-4680-0022}
\affiliation{%
  \institution{Technical University of Munich}
  \country{Germany}
}

\author{Andrés Monroy-Hernández}
\email{andresmh@princeton.edu}
\orcid{0000-0003-4889-9484}
\affiliation{%
  \institution{Princeton University}
  \country{USA}
}

\renewcommand{\shortauthors}{Yuhan Liu et al.}
\begin{abstract}
Beyond the well-known giants like Uber Eats and DoorDash, there are hundreds of independent food delivery platforms in the United States. However, little is known about the sociotechnical landscape of these ``indie'' platforms. In this paper, we analyzed these platforms to understand why they were created, how they operate, and what technologies they use. We collected data on 495 indie platforms and detailed survey responses from 29 platforms. We found that personalized, timely service is a central value of indie platforms, as is a sense of responsibility to the local community they serve. Indie platforms are motivated to provide fair rates for restaurants and couriers. These alternative business practices differentiate them from mainstream platforms. Though indie platforms have plans to expand, a lack of customizability in off-the-shelf software prevents independent platforms from personalizing services for their local communities. We show that these platforms are a widespread and longstanding fixture of the food delivery market. We illustrate the diversity of motivations and values to explain why a one-size-fits-all support is insufficient, and we discuss the siloing of technology that inhibits platforms' growth. Through these insights, we aim to promote future HCI research into the potential development of public-interest technologies for local food delivery. 
\end{abstract}



\received{January 2023}
\received[revised]{July 2023}
\received[accepted]{November 2023}

\maketitle
\input{sections/1_introduction}

\input{sections/2_related_work}
\input{sections/3_method}
\input{sections/4_findings}

\input{sections/5_discussion}
\input{sections/6_limitations}
\begin{acks}
We would like to thank Vibhav Nanda for his support in qualitative coding. We would like to thank our participants for generously sharing their experiences and insights. We also thank Samantha Dalal, Fannie Liu, and other members of the Princeton HCI Lab and Princeton Center for Information Technology Policy for their helpful and thoughtful feedback. Finally, we would like to thank RMDA for sharing their data and offering insightful feedback.
\end{acks}
\bibliographystyle{ACM-Reference-Format}
\bibliography{refbib}

\newpage
\appendix
\input{sections/8_appendix}

\end{document}

%% file: sections/1_introduction.tex
\section{Introduction}
Food delivery platforms provide the logistically challenging service of transporting meal orders from restaurants to customers \cite{kenneyRisePlatformEconomy2016,zwickWelcomeGigEconomy2018}, typically coordinating their work through an array of disparate technical tools. Food delivery platforms have become an invaluable and permanent part of the United States food and hospitality industry ~\cite{anderson2021state}. Three mainstream platforms: \textit{Uber Eats}, \textit{DoorDash}, and \textit{Grubhub} dominate most regions of the United States \cite{ahuja2021ordering}. These mainstream platforms have received considerable attention from scholars and the media, documenting their rise in popularity and impact on local economies ~\cite{veenPlatformCapitalAppetiteControl2020, 
richardsonPlatformsMarketsContingent2020}. In the shadow of these dominating near-monopolies, numerous other food delivery platforms operate across the United States. For years, these platforms have been emerging, growing, failing, and succeeding, while going unnoticed by researchers. The resilience of these platforms, despite the competition of the well-funded players, suggests that they may address unmet needs in the food delivery market. However, little is known about what platforms exist, why they are founded, their operating models, and the regions they serve. 

Within the food delivery landscape in the United States, the widespread, complex phenomenon of independent or ``indie'' platforms is poorly understood. In this work, we use the term ``indie'' platforms to refer to food delivery services that are \textit{not} one of the ``Big Three'' (Uber Eats, Doordash, Grubhub). Scholars have documented the detrimental effects that mainstream platforms have on other participating stakeholders, e.g., couriers ~\cite{saxenaDeliveryAppsAre2019}. However, what role indie platforms play in the food delivery business in the United States remains unexplored. In this paper, we begin to map the landscape of independent restaurant delivery services in the United States. We ask the following research questions:
\begin{enumerate}
\item What are the motivations and values of indie platforms in the United States? 
\item What is the size of their operation?
\item What technical infrastructures do they rely on?
\item What are the challenges they face?

\end{enumerate}

We scraped publicly available online sources to collect data on 495 indie platforms across the United States. We also partnered with the Restaurant Marketing Delivery Association, also known as RMDA, to survey and gather detailed responses from 29 indie platforms to complement insights from the scraped data. In the survey, we asked platforms to describe their organizational history and structure, such as their values, software infrastructure, and vision for the future. 

Through this investigation, we aimed to provide researchers with a description of indie platforms. We learned that indie platforms serve as viable alternatives to the existing mainstream food delivery platforms. However, most of their primary emphasis lies in catering to the local community they engage with. We found that these indie platforms are operated largely on a local level, and they heavily rely on infrastructures designed and built by third-party vendors, which bring obstacles to supporting their communities as they intended. We discuss how existing technical limitations create a disconnect between how indie platforms would like to operate and their current processes, and we propose how social computing researchers can address this gap. We also identified operating challenges indie platforms face, such as a lack of couriers, a low volume of orders, etc.  

In summary, we make two major contributions. First, we map the landscape of independent food delivery platforms in the United States to provide previously unavailable, aggregated insights. Second, we identify opportunities for HCI researchers to support indie platforms in creating sustainable, locally operated alternatives to mainstream platforms.

%% file: sections/2_related_work.tex
\section{Related Work}
The food delivery industry has experienced significant changes in the past decade. Food delivery transitioned from the old model reliant on customers contacting restaurants directly, e.g., calling pizza parlors or Chinese restaurants to get delivery, to online food delivery platforms \cite{hirschberg2016changing}. Online food delivery platforms coordinate the request, preparation, and delivery of food by connecting restaurants, couriers, and customers through their applications. They aggregate supply and demand in the regions they serve~ \cite{kenneyRisePlatformEconomy2016,zwickWelcomeGigEconomy2018}. Several research studies have focused on understanding the dominant online food delivery platforms such as Grubhub, Uber Eats, and Doordash ~\cite{griesbach2019algorithmic,barratt2020m}. Specifically, researchers have investigated their business practices ~\cite{netzer2017urban,gondek2021grubhub}, the technological tools they use~\cite{egg2021online}, and the power relations among platforms, restaurants, consumers, and workers ~\cite{asdecker2020drives}. Nonetheless, little attention has been given to indie food delivery platforms that have different operational, technical, and organizational features~\cite{nosh2023}.  

In this section, we present these two categories of food delivery platforms---mainstream and independent---and give an overview of distinct features that previous research studies have uncovered. First, we discuss the mainstream presence of large platforms, and we describe their business model and operations. We touch on the shortcomings and gaps of mainstream platforms, despite their dominating presence in many regions. Furthermore, we look at how researchers have approached them as a sociotechnical phenomenon. As we demonstrate, little is known about their operations, business values, and technical infrastructure. Our study contributes towards this direction, generating important knowledge on their technical and operational characteristics, the values they possess, as well as how they distinguish themselves from the larger mainstream platforms. 


\subsection{Mainstream Food Delivery Platforms: Features \& Issues}
The United States food delivery landscape in 2022 is dominated by three large platforms: Uber Eats, Doordash, and Grubhub. Together, the ``big three'' dominate the United States with over 90\% collective market share \cite{ahuja2021ordering}. We term these three platforms ``mainstream'' because of their widespread recognition and large market share, and the similarities in their origin, growth, and operations \cite{thompsonHiddenCostFood2020, doornPlatformCapitalismHidden2020, kenneyRisePlatformEconomy2016}. All these three mainstream platforms experienced enormous growth during the past decades by merging with competitors~\cite{zhang2021qualitative}, using exploitative tactics to coerce restaurants to join and stay on the platform~\cite{tkacik2020rescuing} and recruiting a large number of couriers and leveraging state of art algorithms for dispatching to optimize delivery speed and take as many orders as possible ~\cite{gondek2021grubhub}. The large scale and operating model of the mainstream platforms creates a large power asymmetry between them and other stakeholders, especially restaurants and couriers ~\cite{gondek2021grubhub, yao2021together}. 
Given the concentrated power that mainstream platforms possess, recent studies in the CSCW and HCI communities have explored harms associated with them, such as harassment and bias towards vulnerable gig workers, and information asymmetries in algorithm management~\cite{tuco2021food, ma2022brush, kusk2022platform, 10.1145/3563657.3596123}.

Furthermore, researchers have described how the needs of restaurant owners and couriers are not reflected in the design of food delivery platforms ~\cite{zuboff2015big, liBottomUp, Semaan2019,friedmanValueSensitiveDesign2013a,saxenaDeliveryAppsAre2019}. For instance, mainstream platforms’ high commission rates cut into the earnings of already struggling local restaurants. Restaurants are hesitant to raise prices because doing so may drive away price-sensitive consumers~\cite{DeliveryAppsGrubhub2021, thompsonHiddenCostFood2020}. This puts participating restaurants in a situation where they must take a hit on their profit margins to avoid potentially driving away consumers. These negative impacts have driven some restaurants and couriers to explore alternative options for participating in the food delivery platform economy. These alternatives comprise approximately 5-10\% of the market that is not controlled by mainstream platforms \cite{ahuja2021ordering}. In our work, we aim to uncover who these unknown ``other'' players are that continue to exist in a highly competitive market. In the next section, we describe what is known about these smaller platforms and what remains to be investigated.

\subsection{Indie Platforms as an Alternative: An Understudied Sociotechnical Phenomenon}


Similar to other industries where indie is defined by what is not mainstream~\cite{newman2011indie, lipkin2013examining}, any platforms that are not Uber Eats, Doordash, or Grubhub and provide food delivery services for more than one restaurant are broadly categorized as "indie" platforms in this study. No prior research has been conducted to explore indie platforms as a whole. Certain case studies provide us with a solid starting point. Dalal et al. conducted a case study regarding an indie platform in Colorado. The previous study has demonstrated the importance of human intervention when addressing infrastructure breakdown and provides us with insights into this emergent sociotechnical phenomenon from different stakeholders' perspectives~\cite {nosh2023}.  However, little empirical work has been done to detail a fine-grained and integrative understanding of what the indie platform landscape looks like in the United States. We do not know how many of these indie platforms exist, the scale of their operations, the areas they serve, the technological shortcomings they face, and how they sustain and grow in a highly competitive landscape. Thus far, academic research into these new models has been limited, often only focusing on one subset that functions as co-operatives~\cite{bundersFeasibilityPlatformCooperatives2022a,scholz2016and,schor2021after}. 

In this paper, we aim to map the indie platform landscape, uncover aggregated insights into the operations, scales, challenges, and needs of these platforms, and provide direction for CSCW and HCI seeking to better understand indie platforms and the local communities they serve. We thereby initiate a preliminary exploration of the indie online food platform ecosystem., utilizing the online presence of these indie platforms and their organization in associations such as the RMDA. Adopting a mixed-method approach, we combined a large-scale data-driven analysis and small-scale structured surveys; we describe the indie platforms' motivation and values, operational features, technical infrastructure, and challenges in the following sections of this paper. 

%% file: sections/3_method.tex
\section{Method}
We composed a list of 495 U.S.-based indie platforms by aggregating data from news sources, automated web searches, mobile app marketplaces, and collaborations with indie platforms. We then scraped information from their websites and surveyed their operators. We explain these processes below. 

\subsection{Building a List of Indie Platforms}
We gathered a list of 495 indie platforms from the following sources and removed duplicates:

\begin{itemize}
\item \textbf{Seed list.} We obtained a list of 30 platforms from the co-founder of an indie platform we had collaborated with.

\item \textbf{News articles.} We extracted a list of 89 platforms from news articles published from January 2015 to April 2022 from the Media Cloud \footnote{Media Cloud is a constantly growing open-source collection of global news on the open web \cite{roberts2021media} maintained by academic researchers} archive using the keywords \textit{``local food delivery service''} in their API. Two of the authors read the articles and identified the U.S. platforms.
 
\item \textbf{Mobile Apps.}  We identified 259 platforms from the Google Play Store. We searched for apps published by the same publisher of the platforms we had collected already. For example, the app for \textit{Nosh}, a platform in Colorado we had in our list, was published by \textit{Data Dreamers}, a software provider that had published dozens of apps for other platforms. Note that we only collected data from the Google Play Store because the Apple App Store would have produced an incomplete list \footnote{The Apple App Store limits the number of apps uploaded by the same publisher, hence searching by publisher would result in an incomplete list. Publishers seem to circumvent those limitations by creating multiple accounts, which we could not confidently retrieve.}

\item \textbf{Web search.} We retrieved 194 platforms using the Google search API. We searched for local food delivery services in all U.S. towns with a population over 50,000\footnote{We obtained a list of towns from simplemaps.com}. We used the town names to iteratively collect the first 50 search results with a query that  excluded the mainstream platforms and unrelated services, such as meal kit companies and grocery delivery: 
\begin{verbatim}local food delivery city -uber -grubhub -doordash -meal -grocery -farm -courier
\end{verbatim}
These search results still returned irrelevant websites. Members of our research team manually analyzed the websites and extracted names and websites of relevant platforms. 

\item \textbf{Trade organization}. We obtained a list of 230 platforms from the Restaurant Marketing Delivery Association (RMDA), a non-profit that connects restaurant delivery services with one another and organizes a yearly conference. The RMDA's website has an interactive map that includes all their members in the US. 
\end{itemize}



\subsection{Mapping the Geographical Spread of Platforms' Operations}
To map the geographical areas where a platform operated, we collected data on the restaurants available on each platform. We extracted the addresses of 116,855 restaurants from the 328 platforms that were available on the web and their publicly shared lists of cooperating restaurants. Restaurant data included name, address, and contact information. 

Using the topological maps provided by the census bureau \cite{bureau_2022}, we located the boundaries of zip codes in which served restaurants were located for each platform. Then, by using algorithms \ref{algorithm} and \ref{algorithm2}  
(outlined in Appendix \ref{appendix:alg}), we aggregated neighboring zip codes. Algorithm \ref{algorithm} greedily assigned the zip codes of restaurants to clusters based on neighboring adjacency (i.e., shared common borders), while Algorithm \ref{algorithm2} merged clusters that contained common zip codes. In this way, we were able to define how many non-regionally adjacent clusters each platform served. For example, a platform serving a small town would usually serve one cluster, while a platform that would have regional coverage of multiple towns would have more. This clustering approach served as a proxy to indicate the geographical spread of a platform's operations.  

\subsection{Survey Instrument}
To gain further insight into the operations of indie platforms, we developed and deployed an online survey with 31 questions, divided into six sections: Platform Overview, Digital Infrastructure, Restaurant and Customer Relationship, Future Vision, Financial Status, and Open questions. Questions were a mix of multiple-choice and open-text responses. 

We contacted platforms directly using the email listed on their website or via the RMDA mailing list. The email included an overview of our research project, its goals, and a link to the survey. 
We invited owners and managers of delivery platforms to complete a short, 25-minute survey on their experiences with building, maintaining, and growing their organizations. The email stated that survey responses would be kept confidential. Respondents could opt to enter a raffle for a \$100 Amazon gift card. The open-ended responses were coded by two researchers (the first and second authors), following the iterative approach proposed by Braun and Clarke \cite{braun2006using}. The researchers independently coded a subset of the responses and then discussed to reach an agreement on the combined codebooks. These codes were then grouped together into broader themes through a thematic analysis. 

The Princeton University Institutional Review Board approved this work. Survey participants were not obligated to answer every question. The number of responses collected corresponding to each question is listed in Appendix \ref{appendix:count}. 

%% file: sections/4_findings.tex
\section{Findings}
In this section, we answer our four research questions. First, we present several themes that encompass the \textbf{motivations and values} of indie food delivery platforms (RQ1). Next, we describe the \textbf{size of operations} of indie platforms with five different patterns of geographic distribution we observed, along with their weekly order volume (RQ2). We then summarize what we learned about the \textbf{technical infrastructures} used for running their operations (RQ3). Lastly, we discuss the \textbf{challenges} that indie platforms encounter in expanding their presence in a highly competitive market (RQ4). 


\subsection{RQ1: What Are the Motivations and Values of These Platforms?}

The following themes emerged from the websites' analysis and the survey responses:

\subsubsection{Owners belong to the community they serve} Many owners of indie delivery platforms have lived in the community where they operate for many years. These kinship ties with their customers, employees, and local restaurants fostered feelings of responsibility and commitment for platform owners. 

Owners expressed a \textbf{commitment to giving back.} We found that many platforms publicly share how they give back to the communities where they operate. Ways of giving back include donating a portion of profits to local charities and volunteering. For example, owners wanted to allow orders to be rounded up for local charities.

\begin{quote}
    \textit{``During checkout you will be given the option to round up for charity and these proceeds will go directly to Seek and Find Ministries\footnote{A local Christian charity organization} to help support their mission.'' (Swiftly, Illinois)}
\end{quote}

\begin{quote}
    \textit{``At Kershaw2Go, we believe in supporting our community through interactions and donations.  Each quarter, we pick a local non-profit based in Kershaw County and sometimes in surrounding areas and donate towards their cause.'' (Kershaw2Go, South Carolina)}
\end{quote}

We also found that \textbf{owners identify with cause or group.} Many platforms have a strong sense of identity. We found that many platforms proudly promote ownership identity, which includes being owned by women, veterans, or Christians. \textit{Black and Mobile} (Philadelphia) explains how identity shapes their business values: \begin{quote}\textit{``Black and Mobile is the country's first Black Owned food delivery service that exclusively partners with Black-Owned restaurants to give them more exposure and customers.'' (Black and Mobile, Philadelphia)} \end{quote}

In the survey responses, platform owners expressed that \textbf{locally owned and operated businesses have close ties to their community and a personal responsibility to give back.} All respondents described a locally owned and operated business where owners are active within the community. This includes involvement in the daily operations of the business and working ``on the ground,'' as well as community participation outside of their business. One owner explained that they do not believe in correcting errors through giving credits, as is the default with mainstream platforms: \begin{quote}\textit{``We try to go the extra mile for our customers. If something is missing, we always try to go back and get it and not just give a credit.'' (S3)}\end{quote}

Other ways that locally owned businesses distinguish themselves is through serving customers and restaurants that are overlooked by mainstream platforms:
\begin{quote}\textit{``Delivery apps in [location of operation] are restricted to a specific delivery radius and deliver either pizza or fast food.'' (S11)}\end{quote}

Owners also described how they give back to their community:
\begin{quote}\textit{``In 2020 we bought over \$10,000 in Christmas gifts for families in the community'' (S3)}\end{quote}

All indie platform owners detailed the ways in which they serve their communities through active citizenship, beyond simply providing a food delivery service. This commitment to ensuring fair profit-sharing between all parties, quick access to human support when errors occur, and charitable contributions distinguished indie platforms from their competitors. 

\subsubsection{Motivated to provide better price or experience than existing platforms.} Indie platforms believed they offered competitive advantages that existing platforms, including larger competitors, did not. One of these advantages was \textbf{charging low commission rates}. Mainstream platforms charge high commission fees that make it hard for restaurants to sustain their business. Some indie platforms were motivated to offer food delivery at affordable rates: \begin{quote} \textit{``We offer something that apps like Postmates don’t — support and lower commission fees.'' (Foodie Magoo's, Arizona)}\end{quote}
\begin{quote}
    \textit{``We saw local restaurants getting price gouged by the national delivery services, and we wanted to offer a better price and program for restaurants. On that same note, national delivery services take 30\% from restaurants, and 10-15\% from customers, and then pay drivers a low city minimum (\$2.50/order), and they keep the remainder of the profits. We wanted to provide restaurants and drivers both a better opening share of the proceeds.'' (S7)}
\end{quote}

In the survey, platforms described how they charged commission fees. Out of the twenty-six responses we collected, half of the platforms charge fixed percentages ranging from 3\% to 40\%. 27\% of the indie platforms have a sliding model, such as different rates for affiliated and non-affiliated restaurants. One platform charges a flat \$2, while the rest of the 19\% platforms do not charge any commission fees. 

Indie platforms also have the advantage of \textbf{promoting specialty cuisine.} Some indie platforms provide the unique specialization of catering to a certain cuisine or dietary lifestyle. These include platforms that accept only locally owned restaurants, platforms for restaurants that are Halal or Kosher, or platforms that specialize in Asian or African cuisine. Platforms are motivated to elevate their community's food scene, which is consistent with the feelings of connection many platform owners have for their communities. Describing their commitment to staying local, \textit{Chopchop} states: \begin{quote}\textit{``Chris Chandler wants to take food-delivery service out of the hands of the big corporations and provide local service, using a local delivery team, with a local touch'' (Chopchop, Virginia)}  \end{quote} 


In the survey responses, platform owners revealed that \textbf{timely, personalized human involvement differentiates indie platforms.} When asked what differentiates their business from competitors, seventeen of the 29 respondents commented on the personalized service they provide and the ease with which they could be contacted. Seven platforms mentioned keeping profits local and providing better rates for drivers and restaurant owners. Platform owners commented that being local allows for immediate, personalized responses to issues when they arise:

\begin{quote}\textit{``If there's a problem, they call me, the owner of [my platform]. If they need to make a change, they can text, call, or chat with me while I'm picking up an order or eating dinner. I'm not in an office 1200 miles away.'' (S5)}\end{quote}
\begin{quote}
    \textit{``We have people that answer the phone that live in the community so can relate to places and times.'' (S26)}
\end{quote}

This allows them to provide better service, with a platform owner sharing that they outperform the mainstream competitor in their area through faster service, resulting in restaurants leaving the mainstream platform to join their platform. Platform owners compared their business approach to other food delivery business models: \begin{quote}\textit{``(We have) Much higher Service Standards, We are Local, We are personal. Our larger competitors don't do redeliveries when items are missing. We do. Our larger national competitors do not accept cash as a form of payment - We do.'' (S26)}\end{quote}
\begin{quote}
    \textit{``We are customer oriented. We have call center services and resolve issues directly with our customers in real-time. None of these money-losing app competitors can fix a delivery problem -- missing item, wrong item, unsatisfactory item, etc. We can get drivers back at restaurants and do re-trips if needed. Also, unlike the money-losing apps, we can offer instant credits or refunds.'' (S27)}
\end{quote}

\subsubsection{Feel strongly about revitalizing local economies.} Indie platforms were created because founders wanted to support their local economy by creating more job opportunities and keeping profits looping within their local communities. Many platforms advertise that they \textbf{employ locally, or that they partner with local restaurants.} They believe this allows them to provide more personalized, attentive service to customers as they can easily adjust orders. Through their direct connections with restaurants, they can better understand business owners' needs. 

\begin{figure}[htbp]
\centering
\includegraphics[width=0.6\linewidth]{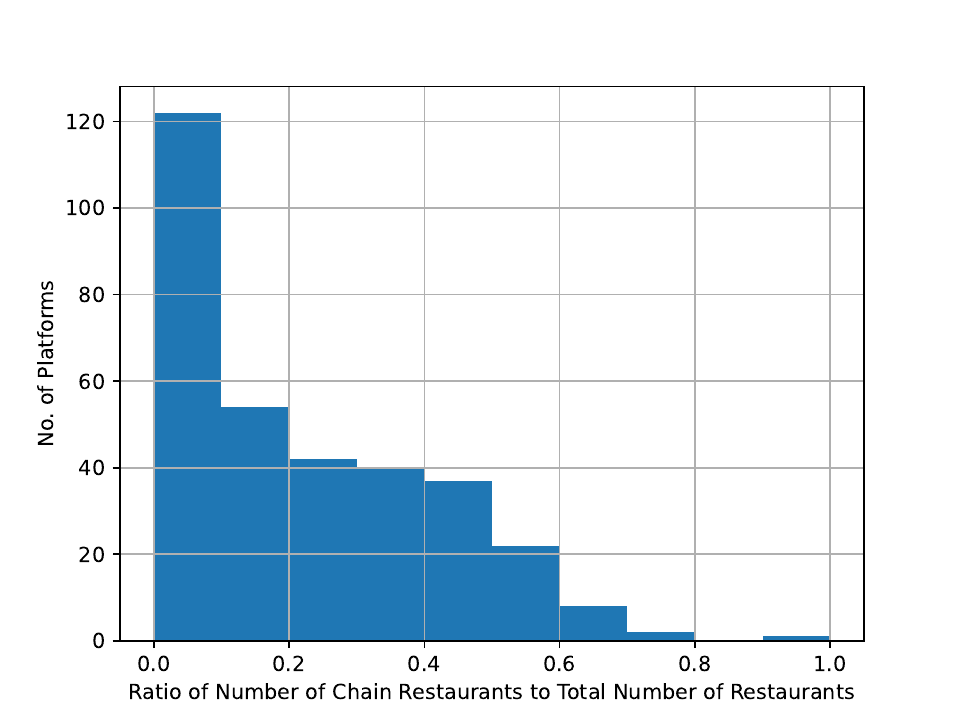}
\caption{Percentage of chain restaurants on indie platform websites}
\label{fig:local_chain}
\end{figure}

In our study, we analyzed indie platforms' collaboration with chain and local restaurants. We used the chain restaurant names listed on Wikipedia to match our data ~\cite{chainwiki} and calculated the percentage of chain restaurants listed on each platform's website. The counts are shown in Figure \ref{fig:local_chain}.  We noticed that more than half of indie platforms(176 out of 326) have less than 20\% chain restaurants listed on their website, while the average percentage of chain restaurants listed on indie platforms is 22.16\%. Note that 50 platforms exclusively partner with local restaurants (no chain restaurant is listed on their website).

Platforms also \textbf{aim to keep money local.} since they are locally owned. This sometimes means that the owners live and work where their platform operates. In other cases, the restaurant owners on a platform also have shares in the platform itself. Through these locally owned and operated business models, indie platforms help to circulate the profits of their success within their local communities. 

The survey responses shed further light on how indie platform owners were \textbf{motivated to create equitable business models in under-served areas.} Four platform owners (13.8\%) saw a business opportunity for food delivery, with three platforms specifying that there were no delivery services in the small towns where they started their businesses. Two platforms were motivated by personal reasons (e.g., leaving a career due to health challenges). Five were motivated to start their business out of a desire to create a locally-owned delivery service. These platforms emphasized care, support, and giving back as central values to their business. One platform owner, who was formerly a driver for a mainstream platform, shared their story:

\begin{quote}\textit{``I was driving for GrubHub, and a local restaurant owner wanted to offer delivery, but the big 4\footnote{The respondent is referring to Uber Eats, Grubhub, Doordash, and Postmates. In this study, we call it the ``big three'' because Postmate was acquired by Uber in 2020. } weren't interested. I started [my platform] to offer local restaurants the same benefit of a shared delivery service that the national chains had.'' (S5)} \end{quote}

Indie platforms expressed their commitment to serving their local community and developing a business model that benefited drivers, restaurant owners, and customers. This motivation often arose after witnessing the impact of mainstream platforms on their community, as described by one platform owner:

\begin{quote}\textit{``We saw local restaurants getting price gouged by the national delivery services, and we wanted to offer a better price and program for restaurants. On that same note, national delivery services take 30\% from restaurants, and 10-15\% from customers, and then pay drivers a low city minimum ([In the city where the platform operates] it is \$2.50/order), and they keep the remainder of the profits. We wanted to provide restaurants and drivers both a better opening share of the proceeds.'' (S7)}\end{quote}

\subsection{RQ2: What Is the Size of Operation of These Platforms?}

\begin{figure}
    \centering
    \subfigure[One Cluster]{
    \begin{minipage}[t]{0.45\linewidth}
    \centering
    \includegraphics[height=1.5in]{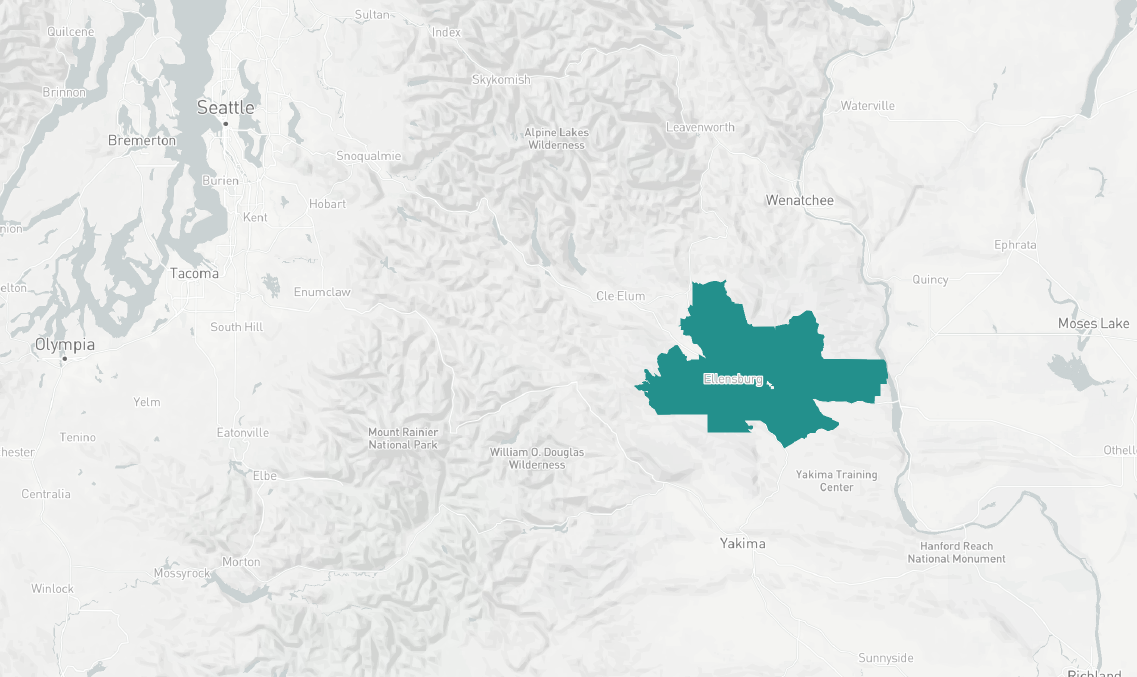}
    \label{subfig:one}
    \end{minipage}%
    }%
    \subfigure[Multiple clusters within a state]{
    \begin{minipage}[t]{0.45\linewidth}
    \centering
    \includegraphics[height=1.5in]{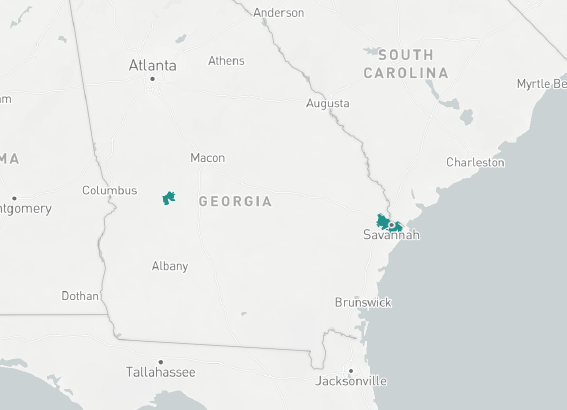}
    \label{subfig:within-state}
    \end{minipage}%
    }%
    \\
    \subfigure[Multiple clusters across neighbouring states]{
    \begin{minipage}[t]{0.45\linewidth}
    \centering
    \includegraphics[height=1.5in]{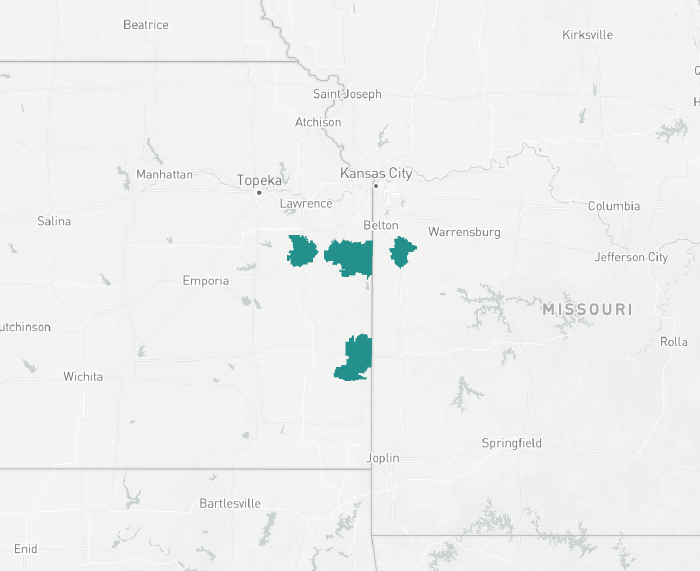}
    \label{subfig:neighboring-states}
    \end{minipage}%
    }%
    \subfigure[Multiple clusters across non-neighbouring states]{
    \begin{minipage}[t]{0.45\linewidth}
    \centering
    \includegraphics[height=1.5in]{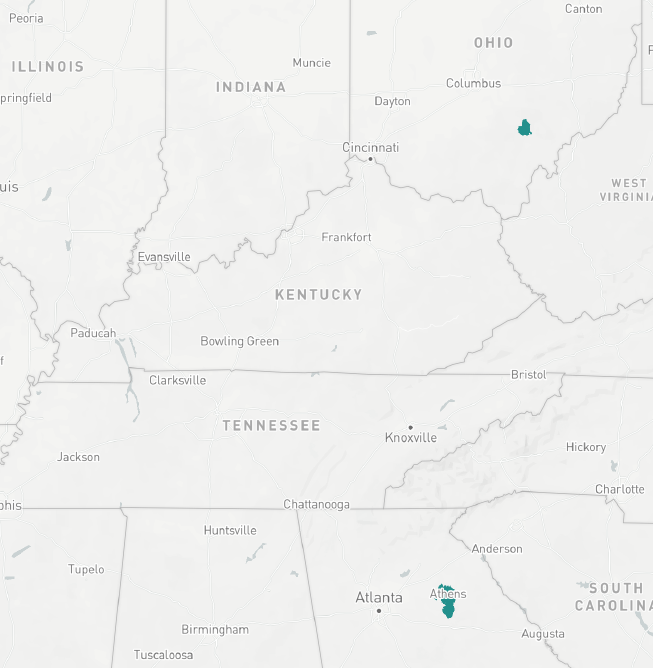}
    \label{subfig:non-neighboring}
    \end{minipage}%
    }%
    \\
    \subfigure[Multiple clusters nationwide]{
    \begin{minipage}[t]{0.4\linewidth}
    \centering
    \includegraphics[height=1.5in]{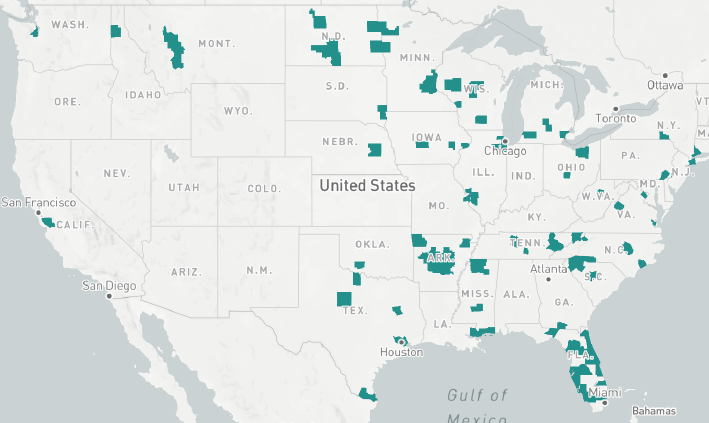}
    \label{subfig:nationwide}
    \end{minipage}%
    }%
    \caption{Examples of indie platforms' operation}
    \label{fig:my_label}
\end{figure}

After uncovering the presence of hundreds of small food delivery platforms, we investigated the size of these operations. Through mapping the restaurant's zip codes, we identified clusters where each platform operates. We found a range in the size of operations, from platforms localized to a single geographic region, to multi-state operations. The variety in geographical dispersion of indie platforms suggests that there are likely diverse motivations, goals, and values that drive their expansion (or lack of it). The three mainstream platforms operate nationwide, with each one offering food delivery in every state across the United States. Though none of the indie platforms operated on the same size as mainstream platforms, we did find that all states but one (Vermont) had an indie platform presence. Below, we list the five categories of operation, from the most geographically concentrated to the most widespread:

\begin{enumerate}
\item \textbf{One cluster.} A platform operating in one cluster serves restaurants that cover one contiguous area. We found 153 (46.65\%) platforms that operate in a single cluster. For example, Figure \ref{subfig:one} is an indie platform operating in Ellensburg, WA.

\item \textbf{Multiple clusters within a state.} Some platforms have restaurants in multiple clusters within a single state. This means the platform serves zones that are geographically separated. We found 97 (29.57\%) platforms that operate in multiple clusters within a state. For example, Figure \ref{subfig:within-state} shows an indie platform operating in Ideal and Savannah, GA.

\item \textbf{Multiple clusters across neighboring states.} Similar to the preceding category, platforms in this category have a presence in geographically distinct zones, though they are in more than one state. However, all states they operate in share a border, suggesting some geographical proximity between clusters. We found 20 (6.1\%) platforms in this category, as shown in. For example, Figure \ref{subfig:neighboring-states} shows an indie platform operating around the border of Kansas and Missouri.

\item \textbf{Multiple clusters across non-neighbouring states.} Platforms in this category also have multiple clusters across states like the prior category. However, a platform in this category operates in states that do not share a border, suggesting a geographical distance between clusters. We found 48 (14.63\%) platforms in this category. For example, Figure \ref{subfig:non-neighboring} shows an indie platform operating in Nelsonville, OH, and Athens, GA.

\item \textbf{Nationwide.} A handful of platforms operate in many states across the United States (i.e., more than 10 states). We categorized these platforms as ``Nationwide'' to distinguish them from the prior category, where a platform operated in no more than 10 states. Ten (3.05\%) platforms operate nationwide as shown in Figure \ref{subfig:nationwide}. 
\end{enumerate}

Besides mapping zip codes to figure out the geographic areas indie platforms cover, we also asked questions regarding the size of operation in our survey to indie platform owners. We received 22 responses to those questions. We found that most of the indie platforms are operating on a small size from the perspective of weekly order volume. In the survey, we asked how many orders the platform received the week before filling out the survey. Excluding one platform that did not receive any orders ``last week,'' 27\% of platforms in our survey indicated receiving less than 100 orders ``last week,'' while 27\% of the platforms received 100-200 orders last week. 32\% platforms received 200-100 orders ``last week'' while 14\% platforms received more than 1,000 orders the week before filling the survey. We found that the order volume is a good indicator to show the size of the platforms as it is highly correlated with other metrics like the number of active customers, the number of couriers doing delivery, and the number of restaurants selling food. For example, all platforms with more than 500 orders have 35-40 couriers for delivery while most of the platforms with less than 200 orders have less than 10 couriers doing delivery the week before the survey. 





\subsection{RQ3: What Technical Infrastructures Do Indie Delivery Services Rely On?}

Through aggregating publicly available data on the app stores, we identified the software and platforms used by indie services. Analysis of survey responses provided descriptive insight into platform owners' concerns with the software they use and software limitations. 

Platforms rely on one, or a combination, of three technical infrastructures for menu listing and ordering: website, iOS app, and Android app. All platforms, except for four, have websites. Those four platforms without websites only have iOS apps. In terms of mobile apps, nearly 70\% of indie platforms have published both iOS and Android apps, while 21\% of have no mobile app. Fourteen platforms only published iOS apps and twenty-seven platforms only published Android apps.

We learned that most platforms did not develop their own software from scratch. Instead, they relied on third-party companies, like Delivery Logic and Data Dreamers. These third-party developers provide a suite of software to handle delivery logistics, including menu-listing software, menu-uploading software for restaurants to manage stock, and software for delivery drivers to handle orders and communicate with dispatchers.

While seven platform owners (26\%) expressed their satisfaction with third-party software, twenty survey responses (74\%) revealed platform owners' dissatisfaction with it. They expressed dissatisfaction with the software limitations, describing a need for greater control and customizability. This included a need for improved customer UI and UX (S4, S6, S15, S18, S21, S22), enhanced communication features with drivers and customers (S3, S5, S10, S13),  integration of front and backend (S7, S20), accurate delivery times (S3), and software that scales (S1).  As reported by one platform owner: 

 \begin{quote}\textit{``(We want) The ability to expand more beyond a local area. The software isn't built for large-scale businesses. There's no way to separate the time zones up for different time zones or any of that.'' (S12)}\end{quote} 
 Most of these respondents who used third-party software feel that the software did not meet their unique needs. For example, Data Dreamers does not allow the platform owner to upload images of food, which creates unnecessary complexity for the platform:   
 
 \begin{quote}\textit{``We can't upload images of food ourselves; the customers can when they get their food, and we can pay them for their images as a rewards program. The images we approve, no matter how good we crop them, end up totally skewed on mobile devices, so it's pretty wacky.'' (S1) }\end{quote} 



These functional limitations impede the quality of services these businesses can provide as well as their growth and ability to compete with larger competition, with one platform owner stating: \begin{quote}\textit{``You would think they [third party software] want the opportunity to help grow some national brands. Ability to change more on the software itself if we wanted to, but there isn't a way. The outdated look of the software is good for those who like more simple software but does look more old school than the larger competition.'' (S1) }\end{quote}  

However, despite the problems platform owners face with third-party software, many of them are dependent on it. Lack of technical skill (twelve platforms) and cost (nine platforms) were the two reasons for selecting an off-the-shelf solution. Hiring engineers to develop and then maintain customized software was cost-prohibitive, as described by two platform owners who explored a custom solution but settled on Data Dreamers and Cartwheel:

\begin{quote}\textit{``Building software is expensive. I talked to a company to build it for us, and it would cost around \$180,000-\$200,000 for them to develop it.'' (S7)}\end{quote}
\begin{quote}
    \textit{``Small companies like mine without any private funding can not afford third-party developer software to create a custom make solution.'' (S29)}
\end{quote}

\subsection{RQ4: What Are the Challenges Indie Platforms Face in Their Current Situation?}

From the survey responses we collected, we learned the challenges indie platforms encountered and how they do or plan to tackle these obstacles. The challenges are closely related to the financial situation and the size of these platforms. Thus we include that information together in this section. 


The most common challenge platforms face is a lack of couriers (17 out of 24), followed by a low volume of orders (10 out of 24), a lack of funding (8 out of 24), a shortage of restaurants (7 out of 24), too many delivery services and not enough customers (5 out of 24), and too many restaurants want to sign up (2 out of 24). One platform faced the challenge of a shortage of employees. No platform indicated they have an oversupply of couriers. 

In order to better understand the challenges, we asked detailed questions about how the platform is doing financially. Fourteen out of twenty-two survey responders mentioned making a profit in the last quarter, one broke even, and seven did not make a profit. 

We also observed that the poor financial situation and lack of couriers are reasons for platforms shutting down in our scraped data. As U.V.E.R, and Somerset Delivery stated on their website: 
\begin{quote}
    \textit{``The service never broke even, required more and more investment to stay alive and the path to cover our costs became too far to reach. After contributing tens of thousands of dollars and several hundred hours of volunteer time, the founders and board members simply cannot justify keeping it alive as a legitimate standalone business.'' (U.V.E.R, New Hampshire)} 
\end{quote}

\begin{quote}
        \textit{``We have spent weeks being unable to keep reliable drivers staffed, which is causing our location to be closed the majority of our available hours. We think it's time to throw in the towel and shut down for good.'' (Somerset Delivery, Kentucky)}
\end{quote}
In fact, we found 128 platforms no longer active \footnote{A platform was classified as active if it had a functioning shopping cart and checkout.} among 495 platforms. Other reasons account for shutdown include acquisition by larger competitors, e.g., \textit{Delivery.com}, \textit{Bite Squad}, or \textit{Grubhub} (resulting in platform websites redirecting to the parent platforms' website), merging with local competitors (resulting in a re-branding), temporarily closed due to technology upgrades, and unknown reasons.

Despite facing similar or different challenges, all platforms had plans to expand in some way. The most common goals in the next year for the platforms we surveyed were:

\begin{enumerate}
\item{Partner with more restaurants (22 platforms).}
\item{Advertise to acquire new customers (20 platforms).} 
\item{Recruit more couriers (20 platforms).}
\item{Expand business to another location (13 platforms).}
\item{Add a service such as catering or alcohol delivery (10 platforms).}
\item{Becoming a delivery provider, i.e., contracting delivery services to third parties (7 platforms).}
\item{Hire more employees (5 platforms).}

\end{enumerate}

%% file: sections/5_discussion.tex
\section{Discussion}

In this section we reflect on indie platforms and what they reveal about their technology use and their technical needs. We organize the Discussion around three takeaways. First, we discuss how our findings suggest that indie platforms are a widespread and longstanding fixture of the United States food delivery industry. However, unlike mainstream platforms, indie platforms operate largely on a local level. This key difference informs our next insight, where we explore why a one-size-fits-all approach to technology cannot address the unique needs of indie platforms. In our third insight, we discuss how food delivery platform technology is heavily siloed and how this siloing may impede indie platforms in performing their business operations and maintaining their values. 


\subsection{Indie Platforms Are a Widespread Fixture of the United States Food Delivery Landscape}

Our analysis indicated that indie food delivery platforms are a significant player in the United States food delivery service. Indie platforms are widespread, existing in all but one state, and have existed for many years, with hundreds still actively operating. We found that most commonly, platforms operate within a single geographical area, followed closely by platforms that operate in multiple geographically separated areas within a state. Most indie platforms are small operations localized within one or a handful of communities, which distinguishes them from mainstream platforms that aggressively pursue expansion into new regions \cite{schor2021after}. 

Mainstream platforms appeal to consumers' demand for low-cost food delivery by charging high commission rates to restaurants rather than the end customer ~\cite{ thompsonHiddenCostFood2020}. However, indie platforms continue to maintain a presence despite the cost-cutting strategies employed by well-funded mainstream platforms \cite{DeliveryCoopsProvide}. The existence of hundreds of indie platforms we discovered suggests that, as similarly reported in prior work, indie platforms position themselves as alternatives to mainstream platforms and are filling gaps in the landscape that are unmet by mainstream platforms ~\cite{atkinsonMoreJobFood2021, DeliveryCoopsProvide}.  In our study, we found that indie platforms adopted a variety of strategies to build presence and trust within the communities where they operate and that it may be these alternative operating models that keep indie platforms in the business. The smaller scale of indie platforms in our study, as well as the community-focused strategies they used, were both key ways in which indie platforms differed from mainstream platforms. Their survival and growth in prominence in such a competitive market are largely attributed to their small-scale, localized nature and human-centered design. In the next insight, we discuss these differences.  

\subsection{One Size Fits All Does Not Cater to the Human Involvement Aspect of Indie Platforms}

We found that indie platforms centered on human involvement throughout their business operations ~\cite{nosh2023}. Owners engaged with day-to-day activities, phone operators were easily accessible to quickly resolve issues, and errors were corrected through personalized service. This person-in-the-loop approach allowed indie platforms to adapt their services and cater to the unique needs of a local community, such as only hiring drivers that have been vetted in person or building a network with local restaurants that mainstream platforms have rejected. Mainstream platforms that aim to minimize reliance on human intervention to reduce operating costs and capture larger market share \cite{hirschberg2016changing, kelsoNowThatDoorDash} are either uninterested in or unable to address such specialized needs on a local scale. The resulting harms of such algorithmic management models are often borne by vulnerable gig workers ~\cite{mohlmann2019people}. The approaches described by indie platform owners in our study aimed to alleviate these harms. 

The focus on mainstream platforms has linked gig work with economies of scale in the public eye, promoting a general perception that gig work must entail an aggressive lowering of costs and increase of production to be profitable \cite{shermanCanUberEver}. Researchers have identified similar issues in India and China ~\cite{seetharaman2021delivery,anjali2022gig,chen2022mixed}. However, our analysis revealed an alternative, viable business model adopted by indie platforms. There is a large digital divide in rural areas ~\cite{duff2021town}. Limited broadband and high-speed internet make human intervention necessary for facilitating digital transactions ~\cite{duff2021town}, and therefore, rural areas are an unappealing market for mainstream platforms that rely on economies of scale \cite{shermanCanUberEver}. Indie platforms in our study filled spaces that are overlooked by larger platforms, such as by serving small rural areas, working with small local restaurants rejected by mainstream platforms, and supporting individuals from under-served communities. From analyzing the motivations of indie platforms, we found that the creation of these platforms was often sparked by identifying a need or gap within their community. As a result, while all indie platforms shared a hands-on philosophy, there was diversity in their specific values, priorities, and business processes. Therefore, a one-size-fits-all approach to technical solutions for food delivery platforms may prevent smaller businesses from engaging with customers, couriers, and restaurants in a way that aligns with their values. We discuss the limitations of existing food delivery technology for indie platforms in the next insight. 



\subsection{Siloing of Technology Impedes Indie Platforms}
At the software level, indie platforms use two suites of software to operate their business: marketplace software for customers, couriers, and restaurants; and dispatch software that handles the delivery logistics. Like the results shown in the previous section, most of the indie platforms in this study rely on third-party software vendors for both types of software. 


We found that this reliance on third-party software poses challenges for indie platforms that differentiate themselves through unique, specialty services. For example, these third-party software vendors use the same templates to build websites and apps for different indie platforms, which makes it hard for platforms to have identifiable impressions through this customer-oriented software. Moreover, there is a lack of customization. Platforms can edit the welcome message, instructions for users, etc, but there is no support for phoning in orders, limitations to setting delivery radius, and insufficient communication for interacting with couriers. However, indie platforms must make do with these off-the-shelf solutions as cost and lack of technical skill prohibit them from developing custom solutions. We discovered a need for technology that empowers indie platforms to customize operational business processes, but the siloing of existing software makes that a challenge. Indie platforms have embraced customers' demands to shift food delivery beyond a simple service. Consumers are willing to pay premiums to enhance the food delivery experience into something fun, and memorable, such as through gourmet meal kits and high-quality packaging  \cite{Gavilan2021innovation}. Furthermore, we noticed the high financial cost from the third-party vendors for indie platforms. They take a fixed margin for each order from indie platforms, ranging from \$0.25-\$0.99, and some charge additional monthly fees and set-up fees, which creates challenges for platforms scaling up.


Here, we propose a concrete path for HCI researchers to develop solutions: a federated, open-sourced framework with a standardized set of protocols. Such design should enable key information revealing like the pricing schema to bring customers more space for comparing the service and quality between indie and mainstream platforms. Furthermore, the framework should include gig worker-centered design to improve the life of couriers \cite{10.1145/3563657.3596123}. Recent literature on Mastodon has already shown the great potential for affordable, federated systems to become the alternative to existing powerful platforms and fill the gap between the demand of users and the design of the platform ~\cite{raman2019challenges, zignani2019footprints}. Similarly, in the federated food delivery network, indie platforms could set up their own instances and customize UI and UX design based on their needs to serve different communities and support different stakeholders. Information and data could flow within the whole network no matter which instance the customer is hosted on since a standardized set of protocols supports the framework. This allows customers to access a broader range of choices from various instances without downloading different apps while preserving their preferred user experience. In addition, The protocol should give due consideration to data portability to ensure that stakeholders, such as restaurant owners, have the ability to control and manage their own data. This aspect can be taken into account along with data signature mechanisms to address the issue of platforms listing restaurant menus without proper consent. A similar design has already been implemented in the ``Bluesky'' \cite{bluesky, blueskyProtocol}. The federated food delivery network also supports in-network collaboration. For example, platforms short of delivery labor could forward the order to other platforms that have extra delivery labor or platforms that are specialized in delivery (like the delivery service provider mentioned in Section 4.4). Such collaboration might be difficult to achieve from a business perspective, but the proposed food delivery network makes the collaboration feasible from a technical perspective. Such solutions could be low-cost and high-customization alternatives to the existing food delivery software indie platforms use.




%% file: sections/6_limitations.tex
\section{Limitations and Future Work}
Our survey consisted of 29 responses, and may not be representative of all indie platform owners. Additionally, the list of indie platforms that we built represents a first attempt at mapping such platforms in the United States. It is likely that we overlooked some platforms in our search, and therefore, this is not an exhaustive investigation into all U.S.-based indie food delivery platforms. To be more specific, our focus is solely on food delivery platforms with web and mobile applications, excluding any other platforms that do not fall within this scope. Moreover, the crawled data is gathered through platform websites without scraping data from any mobile applications. However, only 4 platforms in our study don't have a website, which we believe has a limited impact on the overall results. Furthermore, the identification of indie food delivery platforms heavily relied on keywords and geographic representations. Consequently, there is a possibility that we might have overlooked platforms that do not prominently feature phrases like ``local food delivery'' in their online presence or platforms operating in small cities with populations below 50,000.

Future research could expand the scope of our work by surveying other stakeholders involved in indie food delivery, such as restaurants, couriers, and customers. In our work, we identified the limitations of off-the-shelf solutions. Future work could further investigate the technical needs of indie platforms and the design of open-source software to offer low-cost customizability for indie platforms. Furthermore, studies should focus on the perceptions of consumers towards indie food delivery platforms, in order to understand which factors can contribute to platforms' consumer base growth and sustainable presence in the food delivery space.  

%% file: sections/8_appendix.tex
\section{Clustering algorithms}\label{appendix:alg}
\begin{algorithm}
\caption{ \textit{Greedy restaurant clustering} \\ The algorithm iterates over the zipcodes served by a platform and checks whether they have common boundaries. If yes, it assigns them to the same cluster.}
\label{algorithm}
\begin{algorithmic}[1]
\small
\State $ \texttt{list\_of\_zipcodes} \gets get\_unique\_zipcodes(restaurants_{platform})$
\State $\texttt{list\_of\_clusters} \gets NULL$ 
\For{\texttt{new\_zipcode} in \texttt{list\_of\_zipcodes}}
\If{\texttt{list\_of\_clusters} is \texttt{NULL}}
 \State$\texttt{list\_of\_clusters}.append(\texttt{new\_cluster(new\_zipcode))} $
\Else 
    \State $ \texttt{assign\_zipcode\_to\_cluster} \gets FALSE $ 
    \For{\texttt{cluster} in \texttt{list\_of\_clusters}}
    \For{\texttt{zipcode} in \texttt{cluster}}
        \If{\texttt{zipcode} $\neq$ \texttt{new\_zipcode}}
            \If{is\_adjacent(\texttt{zipcode,new\_zipcode})}
                    \State$\texttt{cluster}.append\texttt{(new\_zipcode)} $
                    \State $ \texttt{assign\_zipcode\_to\_cluster} \gets TRUE $ 
                    \State \textbf{break}
            
    \EndIf
    \Else
      \State $ assign\_zipcode\_to\_cluster \gets TRUE $ 
      \EndIf
       \EndFor
       \EndFor

\If{\texttt{assign\_zipcode\_to\_cluster} is \texttt{FALSE}}
\State$\texttt{list\_of\_clusters}.append(\texttt{new\_cluster(new\_zipcode))} $
    \EndIf
      \EndIf
       \EndFor

\end{algorithmic}
\end{algorithm}
\begin{algorithm}
\caption{\textit{Merge clusters} \\ The algorithm iterates over clusters created by algorithm \ref{algorithm}  and checks whether they share common zipcodes. If yes, it merges them to the same cluster. }
\label{algorithm2}
\begin{algorithmic}[1]
\small
\State $\texttt{list\_of\_merged\_clusters} \gets NULL$ 
    \For{\texttt{cluster} in \texttt{list\_of\_clusters}}
    \If{\texttt{list\_of\_merged\_clusters} is \texttt{NULL}}
\State$\texttt{list\_of\_merged\_clusters}.append(\texttt{cluster)} $
    \Else
        \State $\texttt{Merged} \gets FALSE$ 
        \For{\texttt{new\_cluster} in \texttt{list\_of\_merged\_clusters}}
                \If{have\_common\_zipcodes(\texttt{new\_cluster, cluster})}
                \State $\texttt{new\_cluster} \gets merge(\texttt{new\_cluster, cluster})$ 
                \State $\texttt{Merged} \gets TRUE$ 
        \EndIf
        \EndFor
         \If{\texttt{Merged} is \texttt{FALSE}}
        \State$\texttt{list\_of\_merged\_clusters}.append(\texttt{cluster)} $
        \EndIf
            \EndIf
        \EndFor

\end{algorithmic}

\end{algorithm}

\section{Number of Responses Collected of each question in the Survey}\label{appendix:count}

\begin{table}[H]
  \caption{Number of Responses Collected of each question in the Survey}
  \label{tab:freq}
  \resizebox{\linewidth}{!}{
      \raggedleft
  \begin{tabular}{ll}
    \toprule
    Question& Number of Responses\\
    \midrule
    When was your platform founded? & 29 \\ 
        What motivated you to create your platform? & 29 \\ 
        In what locations does your platform operate? & 29 \\ 
        What distinguishes your platform from other similar services in the same area? & 29 \\ 
        Who built your platform's marketplace (frontend) software?  & 29 \\ 
        Who built your platform's dispatch (backend) software? & 28 \\ 
        What would you like to change about your platform's software, frontend and backend? & 26 \\ 
        You mentioned using self-built software, why did you decide to build the software yourself instead of buying from a third-party company? & 3 \\ 
        You mentioned using software built by another company. Why did you buy that software from a third-party company instead of building it yourself? & 25 \\ 
        What commission fee does your platform charge restaurants? & 26 \\ 
        How does your platform get feedback from restaurants? & 26 \\ 
        How does your platform act on feedback from restaurants, if at all? & 25 \\ 
        What are common frustrations customers encounter when ordering food on your platform? & 25 \\ 
        How does your platform handle those customers' frustrations? & 24 \\ 
        How do customers contact your platform when they have problems or suggestions? & 25 \\ 
        How does your platform respond to customers' problems or suggestions, if at all? & 25 \\ 
        When problems involve costs, who covers those costs?  & 25 \\ 
        Which of the following challenges is your platform facing? & 25 \\ 
        What are the plans for your platform for the next 12 months? & 25 \\ 
        How could policymakers help your platform tackle the challenges it faces, if at all? & 23 \\ 
        How many orders did your platform process last week? & 22 \\ 
        How many restaurants sold food on your platform last week? & 22 \\ 
        How many people ordered food on your platform last week? & 22 \\ 
        How many couriers, e.g., drivers, were delivered for your platform last week? & 22 \\ 
        Financially, how did your platform do last quarter? & 22 \\ 
        What does ``locally owned and operated'' mean to you? & 22 \\ 
        What differentiates a ``locally owned and operated'' restaurant delivery service from other restaurant delivery services? & 20 \\ 
  \bottomrule
\end{tabular}
}
\end{table}